# Bipartite entanglement and hypergraph states

Ri Qu, Juan Wang, Yan-ru Bao, Zong-shang Li, and Yi-ping Ma

*School of Computer Science and Technology, Tianjin University, Tianjin, 300072, China and*

*Tianjin Key Laboratory of Cognitive Computing and Application, Tianjin, 300072, China*

We investigate some properties of multipartite entanglement of hypergraph states in purely hypergraph theoretical terms. We first introduce an approach for computing the concurrence between two specific qubits of a hypergraph state by using the so-called Hamming weights of several special subhypergraphs of the corresponding hypergraph. Then we quantify and characterize bipartite entanglement between each qubit pair of several special hypergraph states in terms of the concurrence obtained by using the above approach. Our main results include that (i) a graph $g$ has a component with the vertex set $\{i,j\}$ if and only if the qubit pair labeled by $\{i,j\}$ of the graph state $|g\rangle$ is entangled; and (ii) each qubit pair of a special hypergraph state is entangled like the generalized W state.

## I. INTRODUCTION

Entanglement is one of the most extraordinary features of quantum theory. It lies at the very heart of quantum information theory [2] and is now regarded as a key physical resource in realizing many quantum information tasks. While the bipartite entanglement is well understood, the ultimate goal to cope with the properties of multipartite entanglement [1] of arbitrary multipartite states is far from being reached. Therefore, several special classes of entangled states have been introduced and identified to be useful for certain tasks. It is well known that *graph states* [3, 4] are an example of these classes. Any graph state can be constructed on the basis of a (simple and undirected) graph. Although graph states can describe a large family of entangled states including *cluster states* [5], *GHZ states*, *stabilizer states* [6], etc., it is clear that they cannot represent all entangled states. To go beyond graph states and still keep the appealing connection to graphs, Ref. [7] introduces an axiomatic framework for mapping graphs to quantum states of a suitable physical system, and extends this framework to directed graphs and weighted graphs. Several classes of multipartite entangled states, such as *qudit graph states* [8], *Gaussian cluster states* [9], *projected entangled pair states* [10], and *quantum random networks* [11], emerge from the axiomatic framework. Moreover, we generalize the above axiomatic framework to encoding hypergraphs into so-called quantum hypergraph states [12]. In [13], we also present an approach for mapping weighted hypergraphs into (up to local unitary transformations) *locally maximally entangable states* [14].

The main aim of this work is to investigate some properties of multipartite entanglement of hypergraph states in purely hypergraph theoretical terms. Several literatures have shown some approaches for this issue. For graph states, Ref. [3] presents various upper and lower bounds to the *Schmidt measure* [15] in graph theoretical terms. For hypergraph states, similar work is done in [12]. Moreover, Ref. [12] qualitatively studies the entanglement structure of hypergraph states in purely hypergraph theoretical terms. Ref. [16] introduces an approach for computing local entropic measure on qubit $t$ of a hypergraph state by using the Hamming weight of the so-called $t$-adjacent subhypergraph. In this paper, we will use the *concurrence* [17] to quantify and characterize the bipartite entanglement between two specific qubits $\{i,j\}$ of a hypergraph state $|g\rangle$ in purely hypergraph theoretical terms. For this, we will present an approach for computing the concurrence between the qubit pair $\{i,j\}$ of the state $|g\rangle$ by using the so-called Hamming weights [16] of several special subhypergraphs of the corresponding hypergraph $g$. Then we will investigate some properties of the entanglement of several special hypergraph states in terms of the concurrence obtained by using the above approach. We will give a sufficiency and necessary condition of two qubits of a graph state being entangled in purely graph theoretical terms. We will also show that a special hypergraph state has the same entangled graph [18] as the generalized W state.

This paper is organized as follows. In Sec. II, we recall notations of hypergraphs, hypergraph states, etc. In Sec. III, we present an approach for computing the concurrence between two specific qubits of a hypergraph state by means of the Hamming weights of some special subhypergraphs. In Sec. IV, we investigate some properties of the entanglement of several special hypergraph states by means of the concurrence. Section V contains our conclusions.

## II. PRELIMINARIES

Formally, a *hypergraph* is a pair $(V,E)$, where $V$ is the set of *vertices*, $E \subseteq \wp(V)$ is the set of *hyperedges* and $\wp(S)$ denotes the power set of the set $S$. The set of all hypergraphs of $n$ vertices is denoted by $\Theta_n$. The *empty hypergraph* is defined as $(V,\varnothing)$. If a hypergraph only contains the *empty hyperedge* $\varnothing$ or one-vertex hyperedges (called *loops*), it is *trivial*. The *rank* of a hypergraph $g$, denoted by $ran(g)$, is the maximum cardinality of a hyperedge in $g$. Moreover, a hypergraph can be depicted by the visual form as shown in Fig. 1. Each vertex is represented as a dot while each hyperedge is represented as a closed curve which encloses the dots corresponding to vertices incident with the hyperedge.

A hypergraph $(V',E')$ is called a *subhypergraph* of $(V,E)$ if $V' \subseteq V$ and $E' \subseteq E$. Let $g=(V,E)$ be a hypergraph. For a vertex $t \in V$ we define the *t-adjacent subhypergraph* $g_t$ of $g$ as $g_t = (V_t, E_t)$ where $V_t = V-\{t\}$ and $E_t = \{e-\{t\} | t \in e \wedge e \in E\}$. For any tow different vertices $i,j \in V$ the $\{i,j\}$-*adjacent subhypergraph* $g_{\{i,j\}}$ and the $(i,j)$-*adjacent subhypergraph* $g_{(i,j)}$ of $g$ are respectively defined as follows: $g_{\{i,j\}} = (V_{\{i,j\}}, E_{\{i,j\}})$ where $V_{\{i,j\}} = V-\{i,j\}$ and $E_{\{i,j\}} = \{e-\{i,j\} | \{i,j\} \subseteq e \wedge e \in E\}$; and $g_{(i,j)} = (V_{(i,j)}, E_{(i,j)})$ where

$V_{(i,j)} = V - \{i,j\}$ and $E_{(i,j)} = \{e - \{i\} \mid i \in e \wedge j \notin e \wedge e \in E\}$.

The vertices incident with the same hyperedge are referred to as being *adjacent*. A sequence of vertices $v_1, v_2, ..., v_p$ such that $v_k$ and $v_{k+1}$ are adjacent for all $1 \leq k \leq p-1$ is called a *path* joining $v_1$ to $v_p$. A hypergraph is *connected* if any two vertices are joined by a path. Otherwise, it is *disconnected*. A *component* of a hypergraph $g$ is a connected subhypergraph contained in no other connected subhypergraph. Moreover, we define the sum of $g = (V, E)$ and $g' = (V', E')$ as $g \Delta g' \equiv (V \cup V', E \Delta E')$ where $E \Delta E'$ denotes the symmetric difference of $E$ and $E'$, that is, $E \Delta E' = E \cup E' - E \cap E'$.

Denote the Pauli matrices by

$$I \equiv \begin{bmatrix} 1 & 0 \\ 0 & 1 \end{bmatrix}, \quad \sigma_x \equiv \begin{bmatrix} 0 & 1 \\ 1 & 0 \end{bmatrix}, \quad \sigma_y \equiv \begin{bmatrix} 0 & -i \\ i & 0 \end{bmatrix}, \quad \sigma_z \equiv \begin{bmatrix} 1 & 0 \\ 0 & -1 \end{bmatrix}. \tag{1}$$

Let $Z_k$ be the $2^k \times 2^k$ diagonal matrix which satisfies

$$(Z_k)_{jj} = \begin{cases} -1 & j = 2^k \\ 1 & others \end{cases} \tag{2}$$

where $k$ is a nonnegative integer. Suppose that $V = [n] \equiv \{1, 2, ..., n\}$ and $e \subseteq V$. Then the *n*-qubit *hyperedge gate* $Z_e$ is defined as $Z_{|e|} \otimes I^{\otimes n-|e|}$ which means that $Z_{|e|}$ acts on the qubits in $e$ while the identity $I$ acts on the rest. An *n*-qubit *hypergraph state* $|g\rangle$ can be constructed by $g = (V, E)$ as follows. Each vertex labels a qubit (associated with a Hilbert space $\mathbb{C}^2$) initialized in $|\phi\rangle = |+\rangle \equiv \frac{1}{\sqrt{2}}(|0\rangle + |1\rangle)$. The state $|g\rangle$ is obtained from the initial state $|+\rangle^{\otimes n}$ by applying the hyperedge gate $Z_e$ for each hyperedge $e \in E$, that is,

$$|g\rangle = \prod_{e \in E} Z_e |+\rangle^{\otimes n}. \tag{3}$$

Thus hypergraph states of $n$ qubits are corresponding to $(\mathbb{C}^2, |+\rangle, \{Z_k \mid 0 \leq k \leq n\})$ by the axiomatic approach shown in [12] while graph states are related with $(\mathbb{C}^2, |+\rangle, Z_2)$ [7,12].

It is known that real equally weighted states [19] are equivalent to hypergraph states [12]. In fact, let $V = [n]$ and define a mapping $c$ on $\wp(V)$ as

$$\forall e \subseteq V, c(e) = \begin{cases} 1 & e = \varnothing \\ \prod_{k \in e} x_k & e \neq \varnothing \end{cases}. \tag{4}$$

Then we can construct a *1-1* mapping $u$ between hypergraphs and Boolean functions which satisfies $\forall g = (V, E)$,

$$u(g)(x_1, x_2, ..., x_n) = \bigoplus_{e \in E} c(e). \tag{5}$$

where $\oplus$ denotes the addition operator over $\mathbb{Z}_2$. Thus we have

$$|g\rangle = \prod_{e \in E} Z_e |+\rangle^{\otimes n} = \frac{1}{\sqrt{2^n}} \sum_{x=0}^{2^n - 1} (-1)^{\bigoplus_{e \in E} c(e)} |x\rangle \equiv |\psi_{u(g)}\rangle \tag{6}$$

where $|\psi_{u(g)}\rangle$ is just the real equally weighted state associate with the Boolean function $u(g)$. Moreover, it is clear that $\forall g, g' \in \Theta_n$,

$$u(g \Delta g') = u(g) \oplus u(g'). \tag{7}$$

It is known that the Hamming weight of a Boolean function $f$ is defined as $|f^{-1}(1)|$ where $|S|$ denotes the cardinality of the set *S*. By (5), we also can define the Hamming weight of a hypergraph *g* with *n* vertices as

$$hw(g) \equiv |f^{-1}(1)| \tag{8}$$

where $f(x_1, x_2, ..., x_n) = u(g)(x_1, x_2, ..., x_n)$. Ref. [16] introduces an approach for calculating the Hamming weight of *g* in purely hypergraph theoretical terms.

## III. CONCURRENCE AND HYPERGRAPH STATES

Concurrence is a famous bipartite entanglement measure. Let $|\phi\rangle$ be a pure state of *n* qubits. The reduced density matrix $\rho_{ij}$ on two different qubits $\{i, j\}$ of $|\phi\rangle$ is defined as $\rho_{ij} \equiv Tr_{\text{all but } \{i,j\}}(|\phi\rangle\langle\phi|)$. One can evaluate the so-called spin-flipped operator defined as

$$\tilde{\rho}_{ij} = (\sigma_y \otimes \sigma_y) \rho_{ij}^* (\sigma_y \otimes \sigma_y) \tag{9}$$

where a star denotes a complex conjugation. Let $\lambda_1$, $\lambda_2$, $\lambda_3$ and $\lambda_4$ be eigenvalues of the matrix $\rho_{ij}\tilde{\rho}_{ij}$ in decreasing order. The concurrence $C_{ij}$ between two qubits $\{i, j\}$ is defined as

$$C_{ij} \equiv \max\left\{0, \sqrt{\lambda_1} - \sqrt{\lambda_2} - \sqrt{\lambda_3} - \sqrt{\lambda_4}\right\} \tag{10}$$

Moreover, it is known that $\rho_{ij}$ is separable or disentangled if and only if $C_{ij} = 0$ [17].

Now let us show how to compute the concurrence between two specific qubits of an $n$-qubit hypergraph state. Let $g = ([n], E)$ be a hypergraph. By (6), the reduced density matrix on two different qubits $\{i, j\}$ of the corresponding hypergraph state $|g\rangle$ can be written into

$$\rho_{ij} = \text{Tr}_{\text{all but } \{i,j\}}(|g\rangle\langle g|) = [a_{rs}]_{4\times 4} \tag{11}$$

where for any $r, s \in \{0, 1, 2, 3\}$

$$a_{rs} = \frac{1}{2^n} \sum_{y=0}^{2^{n-2}-1} (-1)^{u(g)(r,y) \oplus u(g)(s,y)} \tag{12}$$

and

$$u(g)(z, y) \equiv u(g)(x_i, x_j, y) \equiv u(g)(x_1, x_2, ..., x_n). \tag{13}$$

Note that $z = x_i x_j$. For instance, $z = 2$ when $x_i = 1$ and $x_j = 0$. It is similar for $y$. Moreover, it is known that there are four $(n-2)$-valuable Boolean functions $v$, $v'$, $v''$, and $w$ such that

$$u(g)(x_1, x_2, ..., x_n) = x_i x_j v(y) \oplus x_i v'(y) \oplus x_j v''(y) \oplus w(y). \tag{14}$$

Then we can obtain that

$$\forall r \in \{0, 1, 2, 3\}, a_{rr} = \frac{1}{4},$$

$$a_{01} = a_{10} = \frac{1}{2^n} \sum_{y=0}^{2^{n-2}-1} (-1)^{v''(y)}, \quad a_{02} = a_{20} = \frac{1}{2^n} \sum_{y=0}^{2^{n-2}-1} (-1)^{v'(y)}$$

$$a_{03} = a_{30} = \frac{1}{2^n} \sum_{y=0}^{2^{n-2}-1} (-1)^{v(y) \oplus v'(y) \oplus v''(y)}, \quad a_{12} = a_{21} = \frac{1}{2^n} \sum_{y=0}^{2^{n-2}-1} (-1)^{v'(y) \oplus v''(y)}$$

$$a_{13} = a_{31} = \frac{1}{2^n} \sum_{y=0}^{2^{n-2}-1} (-1)^{v(y) \oplus v'(y)}, \quad a_{23} = a_{32} = \frac{1}{2^n} \sum_{y=0}^{2^{n-2}-1} (-1)^{v(y) \oplus v''(y)}. \tag{15}$$

By the definitions of the $\{i, j\}$-adjacent and $(i, j)$-adjacent subhypergraphs, (5), and (14), it is clear that

$$v(y) = u(g_{\{i,j\}}), \quad v'(y) = u(g_{(i,j)}), \quad \text{and} \quad v''(y) = u(g_{(j,i)}). \tag{16}$$

From (7), (8), (15), and (16), we can obtain

$$a_{01} = a_{10} = \frac{1}{4} - \frac{1}{2^{n-1}} hw(g_{(j,i)}), \quad a_{02} = a_{20} = \frac{1}{4} - \frac{1}{2^{n-1}} hw(g_{(i,j)}),$$

$$a_{03} = a_{30} = \frac{1}{4} - \frac{1}{2^{n-1}} hw\left(g_{\{i,j\}} \Delta g_{(i,j)} \Delta g_{(j,i)}\right), \quad a_{12} = a_{21} = \frac{1}{4} - \frac{1}{2^{n-1}} hw\left(g_{(i,j)} \Delta g_{(j,i)}\right),$$

$$a_{13} = a_{31} = \frac{1}{4} - \frac{1}{2^{n-1}} hw\left(g_{\{i,j\}} \Delta g_{(i,j)}\right), \quad a_{23} = a_{32} = \frac{1}{4} - \frac{1}{2^{n-1}} hw\left(g_{\{i,j\}} \Delta g_{(j,i)}\right). \quad (17)$$

Thus it is import for obtaining the reduced density matrix $\rho_{ij}$ to calculate the Hamming weights of the subhypergraphs which occur in (17). It is known that the Hamming weight of a hypergraph can be evaluated by using the approach in [16]. Thus we can obtain the reduced density matrix $\rho_{ij}$ of $|g\rangle$ in purely hypergraph theoretical terms.

In the following, we show how to compute the concurrence of the reduced density matrix $\rho_{ij}$ of the hypergraph state $|g\rangle$. Since all elements of $\rho_{ij}$ in (11) are real, the operator in (9) is equal to $\tilde{\rho}_{ij} = (\sigma_y \otimes \sigma_y) \rho_{ij} (\sigma_y \otimes \sigma_y)$. Clearly, $\lambda_1$, $\lambda_2$, $\lambda_3$ and $\lambda_4$ are corresponding to the squares of eigenvalues of the matrix $\rho_{ij}(\sigma_y \otimes \sigma_y)$ in decreasing order since $\rho_{ij}\tilde{\rho}_{ij} = \left[\rho_{ij}(\sigma_y \otimes \sigma_y)\right]^2$. Thus we can obtain the concurrence of $\rho_{ij}$ according to (10).

## IV. SEVERAL SPECIAL HYPERGRAPH STATES

In this section we discuss some properties of the entanglement of the hypergraph states corresponding to several special hypergraphs by means of the concurrence. These hypergraphs include the hypergraph whose rank is equal to two, the hypergraph $g_n^* = ([n], \{[n]\})$, and so on.

### A. The hypergraph whose rank equals to two

If a hypergraph $g$ is trivial, the concurrence between any two qubits of the hypergraph state $|g\rangle$ is zero since the state $|g\rangle$ is disentangled or fully separable [12]. In the following, we calculate the concurrence between two specific qubits of the hypergraph state corresponding to a hypergraph whose rank equals to two. For convenience, we first define a special function $\varepsilon : \Theta_n \to \{-1, 0, 1\}$ as

$$\forall g = (V, E) \in \Theta_n, \varepsilon(g) = \begin{cases} 1 & E = \varnothing \\ 0 & \text{ran}(g) \geq 1 \\ -1 & E = \{\varnothing\} \end{cases}. \quad (18)$$

Let $g = ([n], E)$ be a hypergraph and $\text{ran}(g) = 2$. Then we can obtain the reduced density matrix $\rho_{ij}$ on two different qubits $i, j \in [n]$ of the hypergraph state $|g\rangle$ as follows.

*Propostion 1.*

(i) If $\varepsilon(g_{\{i,j\}})=1$, then

$$\rho_{ij} = \frac{1}{4}\Big[I\otimes I + \varepsilon(g_{(i,j)})\sigma_x\otimes I + \varepsilon(g_{(j,i)})I\otimes\sigma_x + \varepsilon(g_{(i,j)}\Delta g_{(j,i)})\sigma_x\otimes\sigma_x\Big]. \quad (19)$$

(ii) If $\varepsilon(g_{\{i,j\}})=-1$, then

$$\rho_{ij} = \frac{1}{4}\Big[I\otimes I + \varepsilon(g_{(i,j)})\sigma_x\otimes\sigma_z + \varepsilon(g_{(j,i)})\sigma_z\otimes\sigma_x + \varepsilon(g_{(i,j)}\Delta g_{(j,i)})\sigma_y\otimes\sigma_y\Big]. \quad (20)$$

The above proposition can be described as follows.

*Propostion 1'.* The reduced density matrix $\rho_{ij}$ satifies

$$\rho_{ij} = \frac{1}{4}\Big\{I\otimes I + \varepsilon(g_{(i,j)})\sigma_x\otimes(\sigma_z)^{\delta_{-1,\varepsilon(g_{\{i,j\}})}} + \varepsilon(g_{(j,i)})(\sigma_z)^{\delta_{-1,\varepsilon(g_{\{i,j\}})}}\otimes\sigma_x$$
$$+ \varepsilon(g_{(i,j)}\Delta g_{(j,i)})\Big[\sigma_x\otimes(\sigma_z)^{\delta_{-1,\varepsilon(g_{\{i,j\}})}}\Big]\Big[(\sigma_z)^{\delta_{-1,\varepsilon(g_{\{i,j\}})}}\otimes\sigma_x\Big]\Big\}. \quad (21)$$

where $\delta_{s,t}=1$ if $s=t$; otherwise, $\delta_{s,t}=0$.

*Proof.* It is easy to obtain the reduced density $\rho_{ij}$ according to Sec. III and the properties of the Hamming weights of hypergraphs shown in [16]. For instance, we consider how to obtain $\rho_{ij}$ of $|g\rangle$ when $g_{\{i,j\}}=([n]-\{i,j\},\{\varnothing\})$, $\mathrm{ran}(g_{(i,j)})=1$ and $g_{(j,i)}=([n]-\{i,j\},\varnothing)$. By (18), it is clear that $\varepsilon(g_{\{i,j\}})=-1$, $\varepsilon(g_{(i,j)})=0$ and $\varepsilon(g_{(j,i)})=1$. Since $\mathrm{ran}(g_{(i,j)})=1$ and $g_{(j,i)}=([n]-\{i,j\},\varnothing)$, it is known that $\mathrm{ran}(g_{(i,j)}\Delta g_{(j,i)})=1$. This implies that $\varepsilon(g_{(i,j)}\Delta g_{(j,i)})=0$ by (18). Similarly, we can obtain that $\mathrm{ran}(g_{\{i,j\}}\Delta g_{(i,j)})=1$, $\varepsilon(g_{\{i,j\}}\Delta g_{(j,i)})=-1$, and $\mathrm{ran}(g_{\{i,j\}}\Delta g_{(i,j)}\Delta g_{(j,i)})=1$. According to the proposition 4 in [16] and (17), we can obtain

$$\rho_{ij} = \begin{bmatrix} \frac{1}{4} & \frac{1}{4} & 0 & 0 \\ \frac{1}{4} & \frac{1}{4} & 0 & 0 \\ 0 & 0 & \frac{1}{4} & -\frac{1}{4} \\ 0 & 0 & -\frac{1}{4} & \frac{1}{4} \end{bmatrix} = \frac{1}{4}(I\otimes I + \sigma_z\otimes\sigma_x) \quad (22)$$

which is just the reduced density matrix corresponding to No. 4 shown in the table 2. Moreover, all possible cases of $\rho_{ij}$ of the hypergraph state $|g\rangle$ are shown in the tables 1 and 2. This

implies that (19)-(21) are true. ∎

Table 1. All possible values of the concurrence $C_{ij}$ between two specific qubits $i, j \in [n]$ of the hypergraph state $|g\rangle$ where $\mathrm{ran}(g) = 2$ and $\varepsilon(g_{\{i,j\}}) = 1$.

| No. | $\varepsilon(g_{(i,j)})$ | $\varepsilon(g_{(j,i)})$ | $\varepsilon(g_{(i,j)} \Delta g_{(j,i)})$ | $\rho_{ij}$ | $C_{ij}$ |
|---|---|---|---|---|---|
| 1 | 0 | 0 | 0 | $\frac{1}{4}(I \otimes I)$ | 0 |
| 2 | 0 | 0 | 1 | $\frac{1}{4}(I \otimes I + \sigma_x \otimes \sigma_x)$ | 0 |
| 3 | 0 | 0 | -1 | $\frac{1}{4}(I \otimes I - \sigma_x \otimes \sigma_x)$ | 0 |
| 4 | 0 | 1 | 0 | $\frac{1}{4}(I \otimes I + I \otimes \sigma_x)$ | 0 |
| 5 | 0 | -1 | 0 | $\frac{1}{4}(I \otimes I - I \otimes \sigma_x)$ | 0 |
| 6 | 1 | 0 | 0 | $\frac{1}{4}(I \otimes I + \sigma_x \otimes I)$ | 0 |
| 7 | -1 | 0 | 0 | $\frac{1}{4}(I \otimes I - \sigma_x \otimes I)$ | 0 |
| 8 | 1 | 1 | 1 | $\frac{1}{4}(I + \sigma_x) \otimes (I + \sigma_x)$ | 0 |
| 9 | 1 | -1 | -1 | $\frac{1}{4}(I + \sigma_x) \otimes (I - \sigma_x)$ | 0 |
| 10 | -1 | 1 | -1 | $\frac{1}{4}(I - \sigma_x) \otimes (I + \sigma_x)$ | 0 |
| 11 | -1 | -1 | 1 | $\frac{1}{4}(I - \sigma_x) \otimes (I - \sigma_x)$ | 0 |

According to Sec. III, we can obtain the concurrence $C_{ij}$ by calculating the eigenvalues of $\rho_{ij}(\sigma_y \otimes \sigma_y)$. All possible values of $C_{ij}$ of the hypergraph state $|g\rangle$ are also shown in the tables 1 and 2. Moreover, we can obtain the following proposition by these two tables.

*Proposition 2.* If a hypergraph $g$ has a component whose vertex set is $\{i, j\}$, then $\rho_{ij}$ of the hypergraph state $|g\rangle$ is entangled.

Note the converse of the above proposition is not true since each qubit pair of the hypergraph state $|g_n^*\rangle$ is entangled, which is shown in Sec. IV (B). Now let us discuss some properties of the entanglement of graph states by means of the concurrence. By the above proposition, we can give a sufficiency and necessary condition for two qubits of a graph state being entangled as follows.

*Corollary 3.* A graph $g$ has a component with the vertex set $\{i, j\}$ if and only if $\rho_{ij}$ of the corresponding graph state $|g\rangle$ is entangled.

From the above corollary, we can also obtain the following corollary.

*Corollary 4.* Suppose that $g = ([n], E)$ is a connected graph and $n \geq 3$. Then for any two deferent vertices $i, j \in [n]$ the reduced density matrix $\rho_{ij}$ of the graph state $|g\rangle$ is separable.

Table 2. All possible values of the concurrence $C_{ij}$ between two specific qubits $i, j \in [n]$ of the hypergraph state $|g\rangle$ where $ran(g) = 2$ and $\varepsilon(g_{\{i,j\}}) = -1$.

| No. | $\varepsilon(g_{(i,j)})$ | $\varepsilon(g_{(j,i)})$ | $\varepsilon(g_{(i,j)}\Delta g_{(j,i)})$ | $\rho_{ij}$ | $C_{ij}$ |
|---|---|---|---|---|---|
| 1 | 0 | 0 | 0 | $\frac{1}{4}(I \otimes I)$ | 0 |
| 2 | 0 | 0 | 1 | $\frac{1}{4}(I \otimes I + \sigma_y \otimes \sigma_y)$ | 0 |
| 3 | 0 | 0 | -1 | $\frac{1}{4}(I \otimes I - \sigma_y \otimes \sigma_y)$ | 0 |
| 4 | 0 | 1 | 0 | $\frac{1}{4}(I \otimes I + \sigma_z \otimes \sigma_x)$ | 0 |
| 5 | 0 | -1 | 0 | $\frac{1}{4}(I \otimes I - \sigma_z \otimes \sigma_x)$ | 0 |
| 6 | 1 | 0 | 0 | $\frac{1}{4}(I \otimes I + \sigma_x \otimes \sigma_z)$ | 0 |
| 7 | -1 | 0 | 0 | $\frac{1}{4}(I \otimes I - \sigma_x \otimes \sigma_z)$ | 0 |
| 8 | 1 | 1 | 1 | $\frac{1}{4}(I \otimes I + \sigma_z \otimes \sigma_x)(I \otimes I + \sigma_x \otimes \sigma_z)$ | 1 |
| 9 | 1 | -1 | -1 | $\frac{1}{4}(I \otimes I - \sigma_z \otimes \sigma_x)(I \otimes I + \sigma_x \otimes \sigma_z)$ | 1 |
| 10 | -1 | 1 | -1 | $\frac{1}{4}(I \otimes I + \sigma_z \otimes \sigma_x)(I \otimes I - \sigma_x \otimes \sigma_z)$ | 1 |
| 11 | -1 | -1 | 1 | $\frac{1}{4}(I \otimes I - \sigma_z \otimes \sigma_x)(I \otimes I - \sigma_x \otimes \sigma_z)$ | 1 |

The above corollary has been proved in [20] by using a different approach. It is known that many entanglement criteria (which are shown in [20] and its references) use only bipartite correlations for the entanglement detection. Thus these criteria must fail to recognize in graph states of three or more qubits [20].

Ref. [18] introduces a concept of an entangled graph such that each qubit of a multipartite system is associated with a vertex, while a bipartite entanglement between two specific qubits is represented by an edge between these vertices. For an $n$-qubit state, its entangled graph can visually show how a bipartite entanglement is "distributed" in $n$ qubits. By the corollaries 3 and 4, we can obtain the following proposition.

*Proposition 5.* Suppose that $g = ([n], E)$ is a graph. Then the entangled graph $G = ([n], E_G)$ of the graph state $|g\rangle$ satisfies that each vertex is adjacent with at most one vertex, that is, for each $i \in [n]$

$$|\{j \mid \{i, j\} \in E_G\}| \leq 1. \tag{23}$$

By the above proposition, it is easy to draw all entangled graphs of $n$-qubit graph states. Entangled graphs of three-qubit graph states have been shown in [21]. All Entangled graphs of four-qubit graph states are shown in Fig. 2.

### B. The hypergraph $g_n^* = ([n], \{[n]\})$

In the following, we calculate the concurrence of the reduced density matrix $\rho_{ij}$ on two specific qubits $i, j \in [n]$ of the hypergraph state $|g_n^*\rangle$. According to the proposition 1 in [16] and (17), we can obtain that

$$\rho_{ij} = \begin{bmatrix} \frac{1}{4} & \frac{1}{4} & \frac{1}{4} & \frac{1}{4} - \frac{1}{2^{n-1}} \\ \frac{1}{4} & \frac{1}{4} & \frac{1}{4} & \frac{1}{4} - \frac{1}{2^{n-1}} \\ \frac{1}{4} & \frac{1}{4} & \frac{1}{4} & \frac{1}{4} - \frac{1}{2^{n-1}} \\ \frac{1}{4} - \frac{1}{2^{n-1}} & \frac{1}{4} - \frac{1}{2^{n-1}} & \frac{1}{4} - \frac{1}{2^{n-1}} & \frac{1}{4} \end{bmatrix}. \tag{24}$$

Then we can get

$$\rho_{ij}(\sigma_y \otimes \sigma_y) = \begin{bmatrix} \frac{1}{2^{n-1}} - \frac{1}{4} & \frac{1}{4} & \frac{1}{4} & -\frac{1}{4} \\ \frac{1}{2^{n-1}} - \frac{1}{4} & \frac{1}{4} & \frac{1}{4} & -\frac{1}{4} \\ \frac{1}{2^{n-1}} - \frac{1}{4} & \frac{1}{4} & \frac{1}{4} & -\frac{1}{4} \\ -\frac{1}{4} & \frac{1}{4} - \frac{1}{2^{n-1}} & \frac{1}{4} - \frac{1}{2^{n-1}} & \frac{1}{2^{n-1}} - \frac{1}{4} \end{bmatrix}. \tag{25}$$

Let us calculate four eigenvalues of $\rho_{ij}(\sigma_y \otimes \sigma_y)$. This means that we should solve the equation $\det\left[\rho_{ij}\left(\sigma_y\otimes\sigma_y\right)-\lambda I\otimes I\right]=0$. By computing, we can obtain

$$\det\left[\rho_{ij}\left(\sigma_y\otimes\sigma_y\right)-\lambda I\otimes I\right]=\lambda^2\left(\lambda-\frac{1}{2^{n-1}}-\frac{1}{2^{n/2}}\right)\left(\lambda-\frac{1}{2^{n-1}}+\frac{1}{2^{n/2}}\right) \quad (26)$$

Thus it is known that four eigenvalues of $\rho_{ij}\left(\sigma_y\otimes\sigma_y\right)$ are respectively $\frac{1}{2^{n-1}}+\frac{1}{2^{n/2}}$, $\frac{1}{2^{n-1}}-\frac{1}{2^{n/2}}$, 0 and 0. According to Sec. III, the eigenvalues of $\rho_{ij}\tilde{\rho}_{ij}$, in decreasing order, are $\lambda_1=\left(\frac{1}{2^{n-1}}+\frac{1}{2^{n/2}}\right)^2$, $\lambda_2=\left(\frac{1}{2^{n-1}}-\frac{1}{2^{n/2}}\right)^2$ and $\lambda_3=\lambda_4=0$. From (10), we can obtain that

$$C_{ij}=\frac{2}{2^{n/2}}\neq 0 \quad (27)$$

Therefore we can get the following proposition.

*Proposition 6.* Each qubit pair of the hypergraph state $|g_n^*\rangle$ is entangled.

It is clear that the entangled graph of the state $|g_n^*\rangle$ is a complete graph $K_n$ with $n$ vertices. Ref. [18] shows that the entangled graph of the generalized W state $|W_n\rangle\equiv\frac{1}{\sqrt{n}}\left(|00...01\rangle+|00...10\rangle+...+|10...00\rangle\right)$ is also $K_n$. Thus each qubit pair of the state $|g_n^*\rangle$ is entangled like the state $|W_n\rangle$.

## V. CONCLUSIONS

We first use the Hamming weight of several special subhypergraphs to calculate the concurrence between two specific qubits of a hypergraph state. Then we discuss the properties of the bipartite entanglement of several special hypergraph states by using the concurrence. Our research reveals that the sufficiency and necessary condition of a qubit pair $\{i,j\}$ of a graph state being entangled is that the corresponding graph has a component with the vertex set $\{i,j\}$. Moreover, we also show that the hypergraph state $|g_n^*\rangle$ has the same entangled graph as the generalized W state $|W_n\rangle$. It is interesting that every qubit pair of the state $|g_n^*\rangle$ is entangled like $|W_n\rangle$ while no hypergraph state of $n$ qubits is equivalent to the state $|W_n\rangle$ under local unitraries [13]. This property of bipartite entanglement of the state $|W_n\rangle$ has been used in some quantum information processing tasks. Thus it is helpful for these tasks that the W state $|W_n\rangle$ is replaced to the state $|g_n^*\rangle$ which might be prepared more easily than the state $|W_n\rangle$ in some cases.

## ACKNOWLEDGMENTS

This work is supported by the Chinese National Program on Key Basic Research Project (973 Program, Grant No. 2013CB329304) and the Natural Science Foundation of China (Grant Nos. 61170178, 61272254, 61105072 and 61272265). This work is completed during our academic visiting at Department of Computing, Open University, UK.

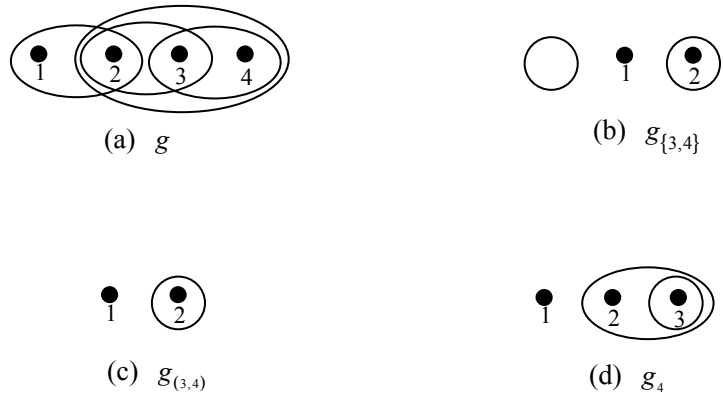

Figure 1. Examples of hypergraphs. The four-vertex hypergraph $g$ in (a) has four hyperedges: $\{1,2\}, \{2,3\}, \{3,4\}$ and $\{2,3,4\}$. In (b), the hypergraph $g_{\{3,4\}}$ with two vertices also has two hyperedges: $\varnothing$ and $\{2\}$. Only one hyperedge, i.e., $\{2\}$, constitutes the hyperedge set of $g_{(3,4)}$ in (c). The three-vertex hypergraph $g_4$ in (d) has two hyperedges: $\{3\}$ and $\{2,3\}$. Moreover, the hypergraphs $g_{\{3,4\}}, g_{(3,4)}$, and $g_4$ are respectively corresponding to the $\{3,4\}$-adjacent, $(3,4)$-adjacent, and 4-adjacent subhypergaphs of $g$. Clearly, the $(4,3)$-adjacent subhypergaph of $g$ is empty.

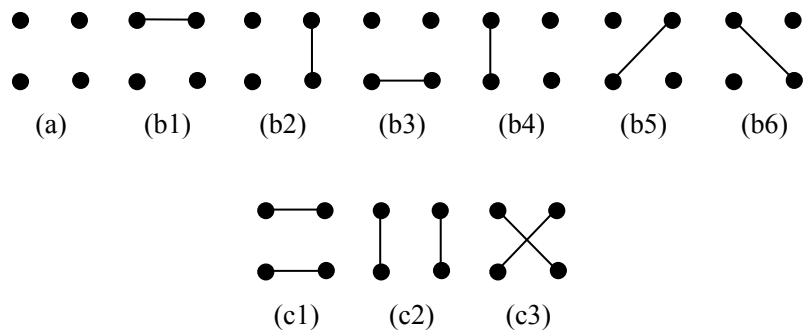

Figure 2. Ten different entangled graphs associated with four-qubit graph states.